\documentclass[12pt]{article}
\usepackage{graphicx}
\usepackage{lscape}
\usepackage{citesort}
\usepackage{amssymb}
\usepackage{appendix}
\usepackage{multirow}

\begin{document}

\def\Tr{\mbox{Tr}}
\def\figt#1#2#3{
        \begin{figure}
        $\left. \right.$
        \vspace*{-2cm}
        \begin{center}
        \includegraphics[width=10cm]{#1}
        \end{center}
        \vspace*{-0.2cm}
        \caption{#3}
        \label{#2}
        \end{figure}
	}
	
\def\figb#1#2#3{
        \begin{figure}
        $\left. \right.$
        \vspace*{-1cm}
        \begin{center}
        \includegraphics[width=10cm]{#1}
        \end{center}
        \vspace*{-0.2cm}
        \caption{#3}
        \label{#2}
        \end{figure}
                }

\def\ds{\displaystyle}
\def\beq{\begin{equation}}
\def\eeq{\end{equation}}
\def\bea{\begin{eqnarray}}
\def\eea{\end{eqnarray}}
\def\beeq{\begin{eqnarray}}
\def\eeeq{\end{eqnarray}}
\def\ve{\vert}
\def\vel{\left|}
\def\ver{\right|}
\def\nnb{\nonumber}
\def\ga{\left(}
\def\dr{\right)}
\def\aga{\left\{}
\def\adr{\right\}}
\def\lla{\left<}
\def\rra{\right>}
\def\rar{\rightarrow}
\def\lrar{\leftrightarrow}  
\def\nnb{\nonumber}
\def\la{\langle}
\def\ra{\rangle}
\def\ba{\begin{array}}
\def\ea{\end{array}}
\def\tr{\mbox{Tr}}
\def\ssp{{\Sigma^{*+}}}
\def\sso{{\Sigma^{*0}}}
\def\ssm{{\Sigma^{*-}}}
\def\xis0{{\Xi^{*0}}}
\def\xism{{\Xi^{*-}}}
\def\qs{\la \bar s s \ra}
\def\qu{\la \bar u u \ra}
\def\qd{\la \bar d d \ra}
\def\qq{\la \bar q q \ra}
\def\GG{\langle g_s^2 G^2 \rangle}
\def\q{\gamma_5 \not\!q}
\def\x{\gamma_5 \not\!x}
\def\g5{\gamma_5}
\def\sb{S_Q^{cf}}
\def\sd{S_d^{be}}
\def\su{S_u^{ad}}
\def\sbp{{S}_Q^{'cf}}
\def\sdp{{S}_d^{'be}}
\def\sup{{S}_u^{'ad}}
\def\ssp{{S}_s^{'??}}

\def\sig{\sigma_{\mu \nu} \gamma_5 p^\mu q^\nu}
\def\fo{f_0(\frac{s_0}{M^2})}
\def\ffi{f_1(\frac{s_0}{M^2})}
\def\fii{f_2(\frac{s_0}{M^2})}
\def\O{{\cal O}}
\def\sl{{\Sigma^0 \Lambda}}
\def\es{\!\!\! &=& \!\!\!}
\def\ap{\!\!\! &\approx& \!\!\!}
\def\ar{&+& \!\!\!}
\def\ek{&-& \!\!\!}
\def\kek{\!\!\!&-& \!\!\!}
\def\cp{&\times& \!\!\!}
\def\se{\!\!\! &\simeq& \!\!\!}
\def\eqv{&\equiv& \!\!\!}
\def\kpm{&\pm& \!\!\!}
\def\kmp{&\mp& \!\!\!}
\def\mcdot{\!\cdot\!}
\def\erar{&\rightarrow&}


\def\simlt{\stackrel{<}{{}_\sim}}
\def\simgt{\stackrel{>}{{}_\sim}}


\title{
         {\Large
    {\bf Masses and Residues of the  Triply Heavy Spin--1/2 Baryons}
         }
      }
\author{\vspace{1cm}\\
{\small  T. M. Aliev$^1$ \thanks {e-mail: taliev@metu.edu.tr}}\,\,
{\small  K. Azizi$^2$ \thanks {e-mail: kazizi@dogus.edu.tr}}\,\,,
{\small M. Savc{\i}$^1$ \thanks
{e-mail: savci@metu.edu.tr}} \\
{\small $^1$ Department of Physics, Middle East Technical University,
06531 Ankara, Turkey}\\
{\small $^2$ Department of Physics, Do\u gu\c s University,
Ac{\i}badem-Kad{\i}k\"oy, 34722 \.{I}stanbul, Turkey}}

\date{}

\begin{titlepage}
\maketitle
\thispagestyle{empty}
\begin{abstract}
We calculate the masses and residues of the triply heavy spin--1/2
baryons using the most general form of their interpolating currents
within the QCD sum rules method. 
We compare the obtained results with the existing theoretical
predictions in the literature.
\end{abstract}
~~~PACS number(s): 11.55.Hx,  14.20.-c, 14.20.Mr, 14.20.Lq
\end{titlepage}

\section{Introduction}

Recently, there have been significant experimental success
on the identification and spectroscopy of the baryons containing
heavy bottom and charm quarks. By this time, all baryons containing
a single charm quark have  been detected as predicted by the
quark model.
The heavy $\Lambda_b$, $\Sigma_b$, $\Xi_b$ and $\Omega_b$ baryons with
spin--1/2 and spin--3/2 $\Sigma_b^\ast$ baryon containing a single
bottom quark have also been discovered (for the current status 
of the heavy flavor baryons see, for example, the review
\cite{Rdhbtt01}).
Recently, CMS Collaboration at CERN  reported the observation of the
spin--3/2 heavy $\Xi_b^\ast$ baryon \cite{Rdhbtt02}. 
SELEX Collaboration announced the first observation
of the doubly heavy spin--1/2  $\Xi_{cc}^+$ baryon with two charm quarks
\cite{Rdhbtt03,Rdhbtt04,Rdhbtt05}. We hope that  the LHCb detector at CERN
will provide us with identification and detection of all doubly heavy and
triply heavy baryons predicted by the quark model.

The experimental progresses on the spectroscopy of the heavy baryons have
stimulated the theoretical studies in this respect. In literature there are
many works on the spectroscopy of the heavy baryons with a single heavy
quark. There are also dozens of works dedicated to the spectroscopy of
the doubly heavy baryons. However, the number of works devoted to the
investigation of the properties of the triply heavy baryons are quite
limited. The spectroscopy of the triply heavy baryons are discussed within
different approaches such as the effective field theory, lattice QCD, QCD bag
model, various quark models, variational approach, hyper central model,
potential model and Regge trajectory ansatz in
\cite{EFT,LQCD1,LQCD2,HaseM,BjorM,JiaM,BeroM,MartM,RobeM,Vijande2004,Patel2009,Llanes-Estrada,XHGuo}.
The masses and residues  of the triply heavy baryons for the Ioffe current
within QCD sum rules method are calculated in \cite{ZhanM,triplyIoffe}.

In the present work we extend our previous studies on the spectroscopy and
mixing angles of the doubly heavy baryons \cite{azizi1,azizi2,azizi3} to the
triply heavy baryons. We calculate the masses and residues of the triply heavy
spin--1/2 baryons using the most general form of their interpolating currents
within the QCD sum rules method. We compare our results with the QCD sum
rule predictions obtained using, the so called, Ioffe current
\cite{ZhanM,triplyIoffe}, as well as with the predictions of other theoretical
approaches
\cite{EFT,LQCD1,LQCD2,HaseM,BjorM,JiaM,BeroM,MartM,RobeM,Vijande2004,Patel2009,Llanes-Estrada,XHGuo}.

The layout  of the article is as follows. In  Section 2, we derive  QCD sum
rules for the masses and residues of the triply heavy spin--1/2 baryons. In
Section 3, we numerically analyze the sum rules for the masses and residues
and find the reliable working regions for the auxiliary parameters that
enter to the sum rules. We compare and discuss our numerical results with the
predictions of the theoretical works existing in the literature.

\section{Masses and residues of the triply heavy spin--1/2 baryons}

In order to obtain the QCD sum rules for the masses and residues of the triply
heavy baryons we start our analysis by considering the correlation function
\bea\label{co}
 \Pi(q)=i \int d^4x e^{i q x} \lla 0 \vel {\cal T}
\big\{\eta_{QQQ'}(x) \bar \eta_{QQQ'}(0)\big\}\ver 0\rra~,
\eea
where $\eta_{QQQ'}$ is the interpolating current for the baryons under
investigation and $q$ is their four-momentum. The most general form of
the interpolating current
for the triply heavy spin--1/2 baryons can  be written as
\bea
\label{cooo}
 \eta_{QQQ'}&=&2\epsilon_{ijk}\Big\{\Big(Q^{iT}CQ^{'j}\Big)\gamma_{5}Q^k+
\beta\Big(Q^{iT}C\gamma_{5}Q^{'j}\Big)Q^k\Big\}~,
\eea
where $i,~j,~k$ are the color indices, $C$ is the charge conjugation operator
and $\beta$ is an arbitrary auxiliary parameter whose working region is to
be determined. The case  $\beta=-1$ in Eq. (\ref{cooo})
corresponding to the Ioffe current is considered
in \cite{ZhanM,triplyIoffe}. The heavy $Q$ and $Q'$ quarks contents of the
triply heavy baryons predicted by the quark model is given in Table 1.
From the current given in Eq. (\ref{cooo}) one can
formally obtain the interpolating current of the proton (neutron) by replacing
$Q\rightarrow u$ and $Q'\rightarrow d$ ($Q\rightarrow d$ and $Q'\rightarrow u$).
\begin{table}[htb]
\begin{center}
\begin{tabular}{|c|c|c|}\hline\hline
             Baryon        &$Q$ &  $Q'$    \\ \hline
      $\Omega_{bbc}$ &  $b$           &$c$     \\ \hline
      $\Omega_{ccb}$ &  $c$           &$b$     \\ \hline
        \end{tabular}
\end{center}
\caption{ The quark contents of the  triply heavy spin--1/2 baryons. }
\end{table}

The  correlation function in Eq. (\ref{co}) can be calculated in two different ways.
On the physical (or phenomenological) side it is calculated in terms of the hadronic
states,
while on the QCD side it is evaluated in terms of quarks and gluons. Matching
these two representations then gives us the QCD sum rules for physical quantities
under consideration. To suppress the contributions of the higher states and continuum
we apply Borel transformation, as well as continuum subtraction to both sides of the
obtained sum rules. 

By saturating the  correlation function on the physical side with a complete set of
hadronic states having the same quantum numbers as the interpolating current and
isolating the ground state baryons, we get
\bea
 \Pi(q)=\frac{\langle 0|\eta_{QQQ'} (0)|B(q)\rangle \langle B(q)|
\bar \eta_{QQQ'} (0)|0\rangle}{q^2-m_B^2}+\cdots~,
\eea
where dots stand for the  contributions coming from the higher states and continuum.
The matrix element of the interpolating current between the vacuum and the
baryonic state is parameterized as,
\bea
 \langle 0|\eta_{QQQ'} (0)|B(q,s)\rangle=\lambda_B u(q,s)~,
\eea
where $\lambda_B$ is the residue of the heavy spin--1/2 baryons and $u(q,s)$ is their
Dirac spinor. By performing summation over the spins of these baryons, we obtain
\bea
 \Pi(q)=\frac{\lambda^{2}_{B}(\rlap/{ q}+m_{B})}{q^2-
m_{B}^2}+\cdots~,
\eea
for the physical side, in which only two independent Lorentz structures $\rlap/{ q}$ and
the identity matrix $I$ survive to be able to calculate the masses and residues of the relevant
baryons. 

On the QCD side, the correlation function is calculated  
using the  operator product expansion (OPE) in deep Euclidean region.
By applying the Wick theorem and contracting out all  quark fields,
we obtain the following expression in terms of the heavy quark propagators:
\bea\label{tree expresion.m}
\Pi(q)&=&4i\epsilon_{ijk}\epsilon_{lmn}\int d^4x e^{i q x}\lla 0 \vel \Big\{-\gamma_{5}
S^{nj}_{Q}S'^{mi}_{Q'}S^{lk}_{Q}\gamma_{5}+
\gamma_{5}S^{nk}_{Q}\gamma_{5}Tr\Big[S^{lj}_{Q}S'^{mi}_{Q'}\Big]
\right.\right.\nnb\\
&+&\left.\left. \beta\Big( -\gamma_{5}S^{nj}_{Q}\gamma_{5}S'^{mi}_{Q'}S^{lk}_{Q}
-S^{nj}_{Q}S'^{mi}_{Q'}\gamma_{5}S^{lk}_{Q}\gamma_{5}
+\gamma_{5}S^{nk}_{Q}Tr\Big[S^{lj}_{Q}\gamma_{5}S'^{mi}_{Q'}\Big]\right.\right.\nnb\\
&+& \left.\left.S^{nk}_{Q}
\gamma_{5}Tr\Big[S^{lj}_{Q}S'^{mi}_{Q'}\gamma_{5}\Big]\Big)+
\beta^2\Big( -S^{nj}_{Q}\gamma_{5}S'^{mi}_{Q'}\gamma_{5}S^{lk}_{Q}+S^{nk}_{Q}
Tr\Big[S^{mi}_{Q'}\gamma_{5}S'^{lj}_{Q}\gamma_{5}\Big]
\Big)
\Big\}\ver 0\rra~,\nnb\\
\eea
where $S'=CS^TC$. 

To proceed on the  QCD side, we write the coefficients of the selected  structures
in terms of the dispersion integral as follows,
\bea
 \Pi_i(q)= \int \frac{\rho_i(s)}{s-q^2} ds~,
\eea
where $\rho_i(s)$ are the  spectral densities and they are determined from the
imaginary parts of the $\Pi_i(q)$ functions. Here $i=1$ and $2$ correspond to
the structures $\rlap/{ q}$ and $I$, respectively. Our main task in the following
is the calculation of these spectral densities. Furthermore, we need  the explicit
expression of the heavy  quark  propagator which is given as, 
\bea
\label{eh32v19}
S_Q(x) &=& {m_Q^2 \over 4 \pi^2} {K_1(m_Q\sqrt{-x^2}) \over \sqrt{-x^2}} -
i {m_Q^2 \rlap/{x} \over 4 \pi^2 x^2} K_2(m_Q\sqrt{-x^2})\nnb \\& -&
ig_s \int {d^4k \over (2\pi)^4} e^{-ikx} \int_0^1
du \Bigg[ {\rlap/k+m_Q \over 2 (m_Q^2-k^2)^2} G^{\mu\nu} (ux)
\sigma_{\mu\nu} +
{u \over m_Q^2-k^2} x_\mu G^{\mu\nu} \gamma_\nu \Bigg]+\cdots~,\nnb \\
\eea
where $K_1$ and $K_2$ are the modified Bessel functions of the second kind.
Substituting this expression of the heavy quark propagator in
Eq. (\ref{tree expresion.m}) and after performing
lengthy calculations  we obtain  the spectral densities
\bea
\rho_1(s) \es \frac{1}{64 \pi^4}\int_{\psi_{min}}^{\psi_{max}}
\int_{\eta_{min}}^{\eta_{max}}d\psi d\eta\Bigg\{-3 \mu_{QQQ'}
\Bigg[-12(-1+\eta)m_Q m_{Q'}(-1+\beta)^2 \nnb\\
\ar \psi^2\eta(3 \mu_{QQQ'}-2s)
\Big[5+\beta (2+5\beta)\Big]+\psi\Bigg(2m^2_Q (-1+\beta)^2 -12m_Q m_{Q'}(-1+\beta^2)\nnb\\
\ar (-1+\eta)\eta(3 \mu_{QQQ'}-2s)\Big[5+\beta (2+5\beta)\Big]\Bigg)\Bigg]\Bigg\}\nnb\\
\ar \frac{\langle g_s^2GG\rangle}{256\pi^4m_Q m_{Q'}}\int_{\psi_{min}}^{\psi_{max}}
\int_{\eta_{min}}^{\eta_{max}}d\psi d\eta
\Bigg\{6(-3+4\psi)(-1+\psi+\eta)m^2_Q(-1+\beta^2)\nnb\\
\ar 6(-3+4\eta)(-1+\psi+\eta)m^2_{Q'}
(-1+\beta^2)+m_Q m_{Q'}\Bigg[48\psi^2(1+\beta^2)+\psi\Big[-63\nnb\\
\ar 68\eta-30\beta+8\eta\beta+
(-63+68\eta)\beta^2\Big]+2(-1+\eta) \Bigg(-3\Big[3+\beta(2+3\beta)\Big]\nnb\\
\ar 2\eta\Big[5+\beta(2+5\beta)\Big]\Bigg)\Bigg]\Bigg\}~, \\ \nnb \\
\rho_2(s) \es \frac{1}{32 \pi^4}\int_{\psi_{min}}^{\psi_{max}}
\int_{\eta_{min}}^{\eta_{max}}d\psi d\eta\Bigg\{3 \mu_{QQQ'}
\Bigg[\eta(-1+\psi+\eta) m_{Q'}(\mu_{QQQ'}-s)(-1+\beta)^2\nnb\\
\ar 6\psi(-1+\psi+\eta) m_{Q}(\mu_{QQQ'}-s)(-1+\beta^2)+m^2_{Q}m_{Q'}
\Big[5+\beta(2+5\beta)\Big]\Bigg]\Bigg\}\nnb\\
\ar \frac{\langle g_s^2GG\rangle}{128\pi^4m_Q m_{Q'}}\int_{\psi_{min}}^{\psi_{max}}
\int_{\eta_{min}}^{\eta_{max}}\frac{d\psi d\eta}{\psi\eta}
\Bigg\{-2(-1+\eta)\eta m_Q m^2_{Q'}(-1+\beta)^2\nnb\\
\ek2\psi^3\eta(-1+\beta)\Big[-9 m_{Q'}(\mu_{QQQ'}-s)(1+\beta)+\eta(2\mu_{QQQ'}-3s)
\Big(m_Q(-1+\beta)\nnb\\
\ar 6m_{Q'}(1+\beta)\Big)\Big]
+\psi m_Q \Bigg(3\eta^3(\mu_{QQQ'}-s)(-1+\beta)^2+2m_Q m_{Q'}(-1+\beta^2)\nnb\\
\ar 3\eta^2\Bigg[-\Big[(\mu_{QQQ'}-s)(-1+\beta)^2\Big]+2m_Q m_{Q'}(-1+\beta^2)+
4m^2_{Q'}(1+\beta^2)\Bigg]\nnb\\
\ar \eta\Bigg[-5m^2_{Q'}(-1+\beta)^2-2m_Q m_{Q'}(-1+\beta^2)+m_Q^2
\Big[5+\beta(2+5\beta)\Big]\Bigg]\Bigg)\nnb\\
\ar \psi^2\Bigg(-4m^2_Q m_{Q'}(-1+\beta^2)
+\eta^2(7\mu_{QQQ'}-9s)(-1+\beta)\Big[m_Q(-1+\beta)+6m_{Q'}(1+\beta)\Big]\nnb\\
\ek 2\eta^3(2 \mu_{QQQ'}-3s)(-1+\beta)\Big[m_Q(-1+\beta)+6m_{Q'}(1+\beta)\Big]
+\eta\Bigg[-18m_{Q'}(\mu_{QQQ'}-s)\nnb\\
\cp (-1+\beta^2)+12m_{Q}m^2_{Q'}(1+\beta^2)-m^3_{Q}\Big[5+\beta(2+5\beta)\Big]
\Bigg]\Bigg)\Bigg\}~,
\eea
where,
\bea
 \mu_{QQQ'} \es \frac{m_Q^2}{1-\psi-\eta}+\frac{m_Q^2}{\eta}+
\frac{m_{Q'}^2}{\psi}-s~,\nnb\\
 \eta_{min} \es \frac{1}{2}\Bigg[1-\psi-\sqrt{(1-\psi)\Big(1-\psi-
\frac{4\psi m_Q^2}{\psi s-m_{Q'}^2}\Big)}~~\Bigg]~,\nnb\\
\eta_{max} \es \frac{1}{2}\Bigg[1-\psi+\sqrt{(1-\psi)\Big(1-\psi-\frac{4\psi m_Q^2}{\psi
s-m_{Q'}^2}\Big)}~~\Bigg]~,\nnb\\
\psi_{min} \es \frac{1}{2s}\Bigg[s+m_{Q'}^2-4m_{Q}^2-
\sqrt{(s+m_{Q'}^2-4m_{Q}^2)^2-4m_{Q'}^2s}~~\Bigg]~,\nnb\\
\psi_{max} \es \frac{1}{2s}\Bigg[s+m_{Q'}^2-4m_{Q}^2+\sqrt{(s+m_{Q'}^2-4m_{Q}^2)^2-
4m_{Q'}^2s}~~\Bigg]~.
\eea

As has already been noted, QCD sum rules for the masses and residues of the
triply heavy baryons can be obtained by
matching the two representations of the correlation function for each structure and
applying the Borel transformation and continuum subtraction to suppress the 
contributions coming from the higher states and continuum, as the result of which we
get,
\bea
\label{sum1}
\lambda^{2}_{B} e^{-m^{2}_{B}/M^2}&=&\int_{s_{min}}^{s_0}
ds \rho_1(s)  e^{-s/M^2}~,\nnb\\
\lambda^{2}_{B} m_{B}e^{-m^{2}_{B}/M^2}&=&
\int_{s_{min}}^{s_0} ds \rho_{2}(s) e^{-s/M^2}~,
\eea
where $M^2$ and $s_0$ are Borel mass parameter and continuum threshold,
respectively, and $s_{min} = (2m_Q+m_{Q'})^2$.
By eliminating the residues from the above equations,
we can calculate the masses of the baryons from either one of the following
expressions,
\bea
\label{sum222}
m^{2}_{B} \es {\ds\int_{s_{min}}^{s_0} ds ~s\rho_i(s)
e^{-s/M^2} \over \ds \int_{s_{min}}^{s_0} ds ~\rho_i(s)
e^{-s/M^2}},~~~~~~~~~i=1~\mbox{or}~2~, \\ \nnb \\
\label{didik}
m_{B} \es {\ds\int_{s_{min}}^{s_0} ds \rho_{2}(s)
e^{-s/M^2}\over \ds \int_{s_{min}}^{s_0} ds \rho_{1}(s)
e^{-s/M^2}}~.
\eea

\section{Numerical results}
Now we are ready to analyze numerically the sum rules obtained in
the previous section and calculate the numerical values of the masses
and residues of the triply heavy spin--1/2  baryons. For this aim we take
the quark masses as their pole values $m_b=(4.8 \pm 0.1)~GeV$ and 
$m_c=(1.46 \pm 0.05)~GeV$ \cite{Colang}, as well as their 
$\overline{MS}$ values   $\bar m_b(\bar m_b)=(4.16 \pm 0.03)~GeV$ and
$\bar m_c(\bar m_c)=(1.28 \pm 0.03)~GeV$ \cite{bikhod}. For the numerical
value of the  gluon condensate we use 
$\langle g_s^2GG \rangle=4 \pi^2 (0.012\pm 0.004)~GeV^4$ \cite{Colang}.

The sum rules obtained in the previous section incorporate also three
auxiliary parameters whose working regions are to be determined.
These parameters are the Borel mass parameter $M^2$, the continuum
threshold $s_0$ and the general parameter $\beta$ enrolled to the general
current of the baryons under consideration. The  working regions of these
parameters are found such that the variations in the values of the masses
and residues are very weak with respect to their running values.

The continuum threshold $s_0$ is not completely arbitrary
and its value is related to the energy of the first excited state.
We do not have adequate information about
the first excited states of the  baryons under consideration,
but our analysis shows that when we choose the  continuum
threshold in the intervals $s_0=(140-148)~GeV^2$ and 
$s_0=(74-81)~GeV^2$,  respectively for the   $\Omega_{bbc}$ and  $\Omega_{ccb}$
baryons, the results  very weakly depend on $s_0$ 
in the case of pole quark masses. While in the case of $\overline{MS}$ values
of the quark masses, the working regions for the continuum threshold
are obtained  as $s_0=(117-125)~GeV^2$ and $s_0=(64-70)~GeV^2$  for
the baryons   $\Omega_{bbc}$ and  $\Omega_{ccb}$, respectively.

Now we proceed to find the working region for the Borel mass parameter $M^2$.
The upper bound on this parameter is found by demanding that  the pole  contribution
 is high compared to the contributions of the continuum and higher states.
This means that the condition,
\bea
\label{nolabel}
{\ds \int_{s_{min}}^{s_0}\ds  \rho(s) e^{-s/M^2} \over
\ds \int_{s_{min}}^\infty \rho(s) e^{-s/M^2}} ~~>~~ 1/2, 
\eea
should be satisfied, which leads to the following upper values for $M^2$:

\bea
\label{e8202}
M_{max}^2 = \left\{ \begin{array}{c}
22~GeV^2,~\mbox{for}~\Omega_{bbc} \\
18~GeV^2,~\mbox{for}~\Omega_{ccb} .
\end{array} \right.
\eea

The lower bound on $M^2$ is calculated requiring that the contribution of the 
perturbative part exceeds   the nonperturbative
contributions. From this restriction we obtain  
\bea
\label{e8202}
M_{min}^2 = \left\{ \begin{array}{c}
12~GeV^2,~\mbox{for}~\Omega_{bbc} \\
9~GeV^2,~\mbox{for}~\Omega_{ccb} .
\end{array} \right.
\eea

Our final task is to determine the working region for the auxiliary parameter
$\beta$. Rather than discussing the variations of the physical observables
with respect to this parameter in the interval $(-\infty,+\infty)$,  we
find it more convenient defining
$\beta=tan\theta$ and look for the variations with respect to  $cos\theta$ in
the interval $-1\le \cos\theta \le 1$. Our numerical results show that in the domains
 $-0.5\le \cos\theta \le-0.9$ and $0.5\le \cos\theta \le 0.9$, the residues depend weakly on $cos\theta$.
Here we should mention that the Ioffe current corresponds to  $cos\theta=-0.71$
and lies inside the reliable region. Note also that, the
masses show a very good stability with respect to $cos\theta$ in the whole
allowed region, whose sum rules are defined
as the ratio of two expressions including  $\beta$ in Eqs. (\ref{sum222}) and (\ref{didik}). 
\begin{table}
\begin{center}
\begin{tabular}{|c|c|c|c|c|c|c|c|}\hline\hline
		      &This work ($\rlap/{ q}$) &This work  ($I$)&  \cite{triplyIoffe}       &\cite{BeroM} &\cite{MartM} &\cite{ZhanM}     &\cite{RobeM}\\ \hline
      $\Omega_{bbc}$ &$11.73\pm0.16$&  $11.71\pm0.16$           &$11.50\pm0.11$    & 11.139      & 11.280      & $10.30\pm0.10$  & 11.535 \\ \hline
      $\Omega_{ccb}$ &$\phantom{1}8.50\pm0.12$&  $\phantom{1}8.48\pm0.12$
      &$\phantom{1}8.23\pm0.13$     & \phantom{1}7.984       & \phantom{1}8.018       &
      $\phantom{1}7.41\pm0.13$   & \phantom{1}8.245  \\ \hline
      $\overline{\Omega}_{bbc}$ &$10.59\pm0.14$&  $10.56\pm0.14$           &$10.47\pm0.12$    & -     & -     & - & - \\ \hline
      $\overline{\Omega}_{ccb}$ &$\phantom{1}7.79\pm0.11$&  $\phantom{1}7.74\pm0.11$
      &$\phantom{1}7.61\pm0.13$     & -      & -      & -   & -  \\ \hline
        \end{tabular}
\end{center}
\caption{ The masses of the triply heavy spin--1/2 baryons (in units of
$GeV$). For the  baryons with over-line,
the $\overline{MS}$ values of the quark masses 
are used. }
\end{table}

\begin{table}
\begin{center}
\begin{tabular}{|c|c|c|c|}\hline\hline
                     &This  work ($\rlap/{ q}$) &This work  ($I$) & \cite{triplyIoffe}       \\ \hline
      $\Omega_{bbc}$ &  $0.53\pm0.17$  &   $0.45\pm0.15$      &$0.68\pm0.15$     \\ \hline
      $\Omega_{ccb}$ &  $0.38\pm0.13$    &   $0.30\pm0.10$    &$0.47\pm0.10$     \\ \hline
      $\overline{\Omega}_{bbc}$ &  $0.85\pm0.28$  &   $0.65\pm0.22$      &$0.68\pm0.15$     \\ \hline
      $\overline{\Omega}_{ccb}$ &  $0.56\pm0.18$    &    $0.38\pm0.13$    &$0.47\pm0.10$     \\ \hline
        \end{tabular}
\end{center}
\caption{ The residues of the triply heavy spin--1/2 baryons (in units of
$GeV^3$).
For the baryons with over-line, the $\overline{MS}$ values of the quark masses 
are used. }
\end{table}

Considering the working regions of the auxiliary parameters we obtain the
numerical values for the masses and residues of the triply heavy spin--1/2
baryons as presented in Tables 2 and 3 for both
structures. For comparison we also 
present the numerical predictions of  other theoretical approaches such as
the modified bag model \cite{BeroM}, relativistic quark model
\cite{MartM}, non-relativistic quark model \cite{RobeM}
and QCD sum rules for the Ioffe current \cite{ZhanM,triplyIoffe} in the same Tables. As far as
the masses are considered, our central value results are slightly higher
than the other predictions. The closest results to our predictions are the
results of
the non-relativistic quark model \cite{RobeM} and QCD sum rules with
the Ioffe current \cite{triplyIoffe}, respectively. The lower predictions
for the masses belong, respectively, to
QCD sum rules with the Ioffe current \cite{ZhanM} and the modified bag
model \cite{BeroM}.  From Table 2 we  see that the two structures in our
case give approximately the same results. This Table also shows that 
the results depend on the quarks masses considerably and change (9-10)\% when
one proceeds from the pole to the $\overline{MS}$ scheme mass parameters. Here, we
should mention that considering  Eq. (\ref{didik}) does not
affect considerably the results of masses presented in Table 2.

In the case of the residues, in contrast to the predictions given in
\cite{triplyIoffe}, our
results depend on the quark masses in such a way that when we switch
from the pole to the $\overline{MS}$ scheme quark mass parameters,
our results change (21-37)\%. This is an expected
result since the residues depend more on quark masses in comparison with
the baryon masses. From Table 3 it is also clear that the results depend
on the choice of the structure. The structure $I$ gives the results
(15-30)\% lower compared to those of the structure $\rlap/{ q}$.
In the case of the pole masses of the quarks, our predictions on the
residues are considerably smaller in comparison with those of the
\cite{triplyIoffe}. The maximum difference between two works is observed for
the residue of the $\Omega_{ccb}$ baryon obtained from the $I$ structure,
which is approximately 36\%.
For the residues in $\overline{MS}$ scheme, our results are very close to
those of the \cite{triplyIoffe} for the structure $I$ and 
$\overline{\Omega}_{bbc}$, while the maximum difference
of 23\%   between predictions of two studies belongs to
the $\overline{\Omega}_{ccb}$ baryon and also the structure $I$.

In conclusion, we calculated the masses and residues of the triply
heavy spin--1/2 baryons using  the most general form of their interpolating
currents in the framework of QCD sum rules.
We found the reliable working regions of the auxiliary parameters
entered to the mass and residue calculations.
Our predictions on the masses are slightly higher than the predictions
of the other approaches such as, the modified bag model,
relativistic and non-relativistic
quark models as well as QCD sum rules for 
the Ioffe current. The predictions for the residues we obtained are
considerably different compared to the
present predictions of the QCD sum rules for the Ioffe current. We hope that
the LHC at CERN
will provide opportunity  to experimental study of these baryons
in near future.

\end{document}